\documentclass[osajnl,twocolumn,showpacs,superscriptaddress,10pt]{revtex4-1}

\usepackage{amsmath,amssymb,graphicx}

\begin{document}

\title{Quantum frequency conversion of vacuum squeezed light to bright tunable blue squeezed light and higher-order spatial modes}

\author{Hugo Kerdoncuf}
\affiliation{Danish Fundamental Metrology, Kogle Alle 5, 2970 H{\o}rsholm,  Denmark}
\affiliation{Current workplace: NKT Photonics A/S, Blokken 84, 3460 Birkerød, Denmark}

\author{Jesper B. Christensen}
\affiliation{Danish Fundamental Metrology, Kogle Alle 5, 2970 H{\o}rsholm,  Denmark}

\author{Mikael Lassen}\email{Corresponding author: ml@dfm.dk}
\affiliation{Danish Fundamental Metrology, Kogle Alle 5, 2970 H{\o}rsholm,  Denmark}

\begin{abstract}
Quantum frequency conversion, the process of shifting the frequency of an optical quantum state while preserving quantum coherence, can be used to produce non-classical light at otherwise unapproachable wavelengths. We present experimental results based on highly efficient sum-frequency generation (SFG) between a vacuum squeezed state at 1064 nm and a tunable pump source at 850 nm $\pm$ 50 nm for the generation of bright squeezed light at 472~nm $\pm$ 4~nm, currently limited by the phase-matching of the used nonlinear crystal. We demonstrate that the SFG process conserves part of the quantum coherence as a 4.2($\pm0.2$)~dB 1064 nm vacuum squeezed state is converted to a 1.6($\pm$0.2)~dB tunable bright blue squeezed state. We furthermore demonstrate simultaneous frequency- and spatial-mode conversion of the 1064-nm vacuum squeezed state, and measure 1.1($\pm$0.2)~dB and 0.4($\pm$0.2)~dB of squeezing in the TEM$_{01}$ and TEM$_{02}$ modes, respectively. With further development, we foresee that the source may find use within fields such as sensing, metrology, spectroscopy, and imaging.  
\end{abstract}

\maketitle

%%%%%%%%%%%%%%%%%%%%%%%%%%  body  %%%%%%%%%%%%%%%%%%%%%%%%%%
\section{Introduction}
Optical quantum sensing exploits the unique quantum correlations of non-classical light to enhance the detection of physical parameters beyond classical means \cite{Giovannetti2004,Pirandola2018,Lawrie2019}. While several different classes of quantum states may provide such a quantum advantage, it is only the squeezed states of light, which due to their relative generation simplicity, brightness, and robustness, have been demonstrated to be applicable in practice \cite{Walls1983,Andersen2016}. Squeezed states of light therefore represent a particularly useful resource to enhance the detection of physical parameters beyond classical means and can be useful for measurements of extremely weak signals, with the crowning example being the detection of gravitational waves \cite{Ligo2011,Mehmet2018}. Recent examples also include the use of squeezed light for bio-optical measurements for improving the signal-to-noise ratio of weak bio-optical signatures \cite{Taylor2013,Andrade2020,Casacio2021}. In general, tunable squeezed states at visible wavelengths are of great interest and is envisioned to have a large variety of applications in fundamental research and medical diagnosis \cite{Ye2008,Yun2017,Allgaier2017,Braunstein2005}. Visible light provides tighter optical focusing capabilities than the infrared counterpart, resulting in superior imaging and spectroscopic potential. Moreover, interferometric measurement sensitivity scales inversely with wavelength, making the blue- and UV spectral regions particularly compelling from a sensing- and metrological perspective.  \cite{Polzik1992,Bell1997,LassenPRL2007,Tsuchida1995,Baune2015,Lassen2007,Janousek2009}. There is therefore an unmet need for novel types of squeezing sources that covers a big wavelength range to provide a quantum bridge between different wavelengths and can operate on any higher-order spatial modes.

So far the most successful method for the generation of strongly squeezed states of light is using continuous wave (CW) optical parametric amplifiers (OPAs) pumped by a strong second-harmonic (SH) pump field \cite{Takeno2007,Vahlbruch2016}. However, the OPA technology has primarily been implemented in the near-infrared (NIR) wavelength region as the generation of squeezed states at visible and, in particular blue, wavelengths using OPAs is hampered by technical difficulties. Most notably, intense SH fields in the ultraviolet wavelength range, used for pumping the OPA, leads to undesired absorption effects and potentially photorefractive damage of the nonlinear crystal \cite{Boulanger1999,Hirohashi2007}. Fortunately, highly efficient nonlinear frequency-conversion preserves the quantum coherence such as those observed for squeezed states of light. This nonlinear process, named quantum frequency conversion (QFC) \cite{Kumar1990,Huang1992,Mckinstrie2005}, may therefore be used to bridge dissimilar quantum systems that require different optical career frequencies or optical bandwidths \cite{Tanzilli2005,Mcguinness10,Allgaier2017}. Moreover, QFC can be exploited to produce quantum-correlated states of light, at optical frequencies which are otherwise challenging to access. For example, it has been demonstrated that a squeezed vacuum can be converted from the NIR at 1550~nm to the green region at 532~nm through sum-frequency generation (SFG) driven by a strong coherent pump at 810 nm \cite{Vollmer2014,Baune2015}. By this principle, SFG enables the production of squeezed quantum states of light in the complete ultraviolet to visible range by using the coherent pump laser for wavelength selectivity --- ultimately realizing a source of tunable squeezed-light at wavelengths that are otherwise difficult to access \cite{Bell1997,Tsuchida1995,Baune2015,Allgaier2017,Mimoun2008,Hansen2015,Kerdoncuff2020}.

\begin{figure}[htbp]
\centering
\fbox{\includegraphics[width=0.9\linewidth]{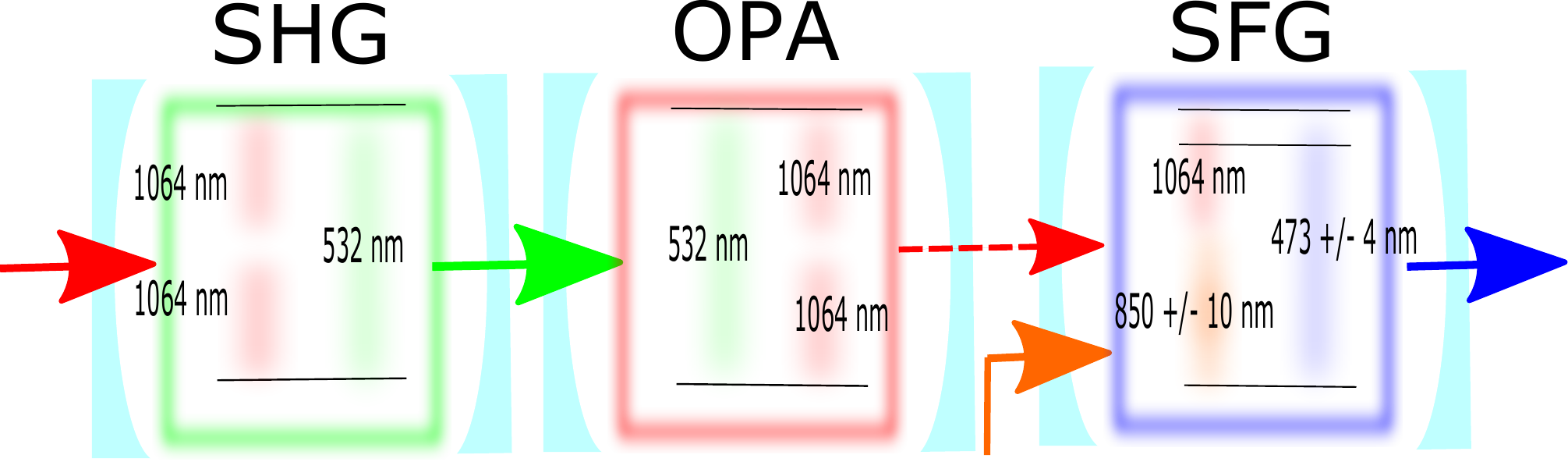}}
\caption{Block diagram of the second-order nonlinear interactions used to generate squeezed blue light, where the energy levels of the nonlinear processes also are illustrated. Second-harmonic generation (SHG), optical parametric amplification (OPA), and sum-frequency generation (SFG). All second-order $(\chi^{(2)})$ nonlinear interactions are cavity-enhanced in the experiments. }
\label{fig0}
\end{figure}

Here we present the experimental realization of a QFC system based on highly efficient SFG in a double resonant cavity, where a tunable pump beam at 850~nm $\pm$ 50 nm and a vacuum squeezed seed beam at 1064~nm are mixed using a periodically poled (PP)KTP nonlinear crystal to produce wavelength-tunable bright squeezed light in the blue spectral region, see Fig.~\ref{fig0} for the principle outline of the experiment and the cascaded nonlinear interactions. A single-resonant degenerate OPA is used to generate a 4.2($\pm$0.2) dB squeezed vacuum state at 1064~nm, which upon frequency conversion to $\sim$473 nm, is converted to a bright blue squeezed state with a squeezing strength of 1.6($\pm$0.2)~dB. We further demonstrate the ability to simultaneously convert the spatial mode of the squeezed light \cite{LassenPRL2007}, measuring preserved squeezing levels of 1.1($\pm$0.2)~dB and 0.4($\pm$0.2)~dB in a TEM$_{01}$ and TEM$_{02}$ mode, respectively.  To our knowledge, this is the first demonstration of tunable squeezed light in higher order spatial modes using an SFG stage seeded with 1064 nm vacuum squeezed states.

\section{Experimental setup}

\begin{figure}[htbp]
\centering
\fbox{\includegraphics[width=0.9\linewidth]{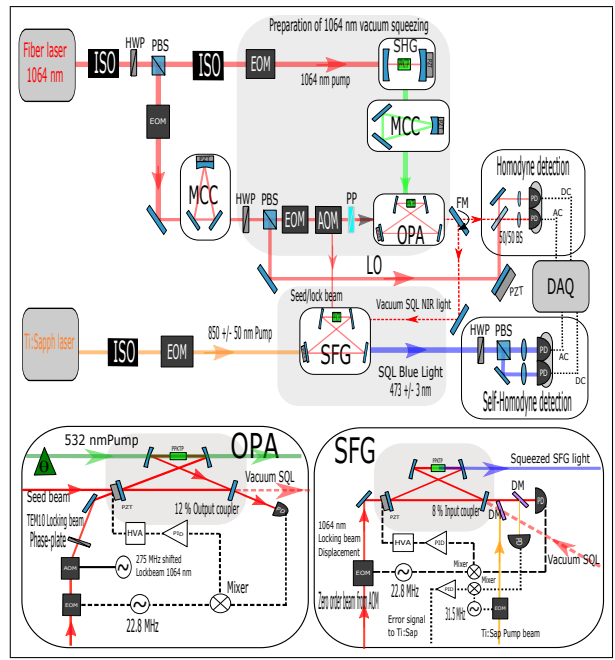}}
\caption{Block diagram of the experimental setup with the main components. The two lasers used are a a 1064~nm CW fiber laser and a tunable CW Ti:sapphire laser. The experimental setup involves three nonlinear optical interactions (SHG, OPA and SFG) and two mode-cleaner cavities. Squeezed vacuum states (illustrated by dashed lines) at 1064 nm were produced in a degenerate OPA and up-converted in the SFG double resonant cavity to 473 nm $\pm$ 4 nm. The reflection and the transmission from the cavities are used for PDH locking. The generated quantum states at 1064~nm and 472~nm $\pm$ 4~nm are detected with homodyne detection. ISO: optical isolator. MCC: mode-cleaner cavity. SHG: second-harmonic generation. OPA: optical parametric amplification. SFG: sum-frequency generation. EOM: electro-optical modulator. HWP: half-wave plate, PBS: polarized beam splitter. PZT: piezo-electrical element. PP: phase plate. FM: flip mirror.}
\label{fig1}
\end{figure}

The schematic of our setup is depicted in Fig.~\ref{fig1}. The two laser sources used for the experiment are a 1064~nm CW fiber laser with a 10~kHz linewidth and up to 10~W output power, and a CW Ti:sapphire laser tunable in the range from 800~nm to 900~nm with 75~kHz linewidth and delivering a maximum power of 260~mW to the SFG cavity. The setup consist of two mode-cleaner cavities (MCC), a linear second harmonic generation (SHG) cavity (1064 nm to 532 nm), an OPA cavity (pump at 532~nm generating vacuum squeezing at 1064~nm), and a double resonant (DR) SFG cavity (for mixing 1064 nm and 850 $\pm$ 4 nm light). To keep the seed, displacement, and pump beams on resonance with the cavities, we apply Pound-Drever-Hall (PDH) locking methods by demodulating the measurements of the cavity reflections and transmissions \cite{PDH}. Phase modulations via electro-optical modulators (EOM) at different frequencies, above the bandwidths of the different cavities, are chosen so that potential beating signals within the measurement bandwidth are minimized. Details about the locking schemes of the OPA and SFG cavity can also be  seen in figure\ref{fig1}. We use homodyne detection- and self-homodyne detection schemes for measuring the generated squeezed states at 1064~nm and 473 $\pm$ 3~nm, respectively. Part of the 1064 nm beam is first directed to the SHG in order to generate a bright 532~nm pump for the OPA. The SHG cavity is a standing-wave cavity consisting of two curved mirrors of 50~mm radius of curvature and a 20 mm long periodically poled KTP (PPKTP) crystal placed in the middle. The input coupling mirror has a transmittance of $T = 8~\%$ at the 1064~nm pump wavelength and more than $T= 95~\%$ for the 532 nm second harmonic (SH) wavelength, and the back mirror is highly reflective (HR) for both 1064 nm and 532 nm. In this cavity, more than 600 mW of 532-nm light is produced through SHG for pumping the OPA. The remaining part of the 1064-nm beam and the 532-nm beam are both directed through MCCs, which filters the intensity noise of the laser above the bandwidth of the MCC and spatially filters the modes. Bandwidths of 1.5~MHz are measured for both cavities and a transmission greater than 85$\%$ is obtained for the TEM$_{00}$ mode. 

\section{Preparation of 1064 nm vacuum squeezing}

\begin{figure}[htbp]
\centering
\fbox{\includegraphics[width=0.9\linewidth]{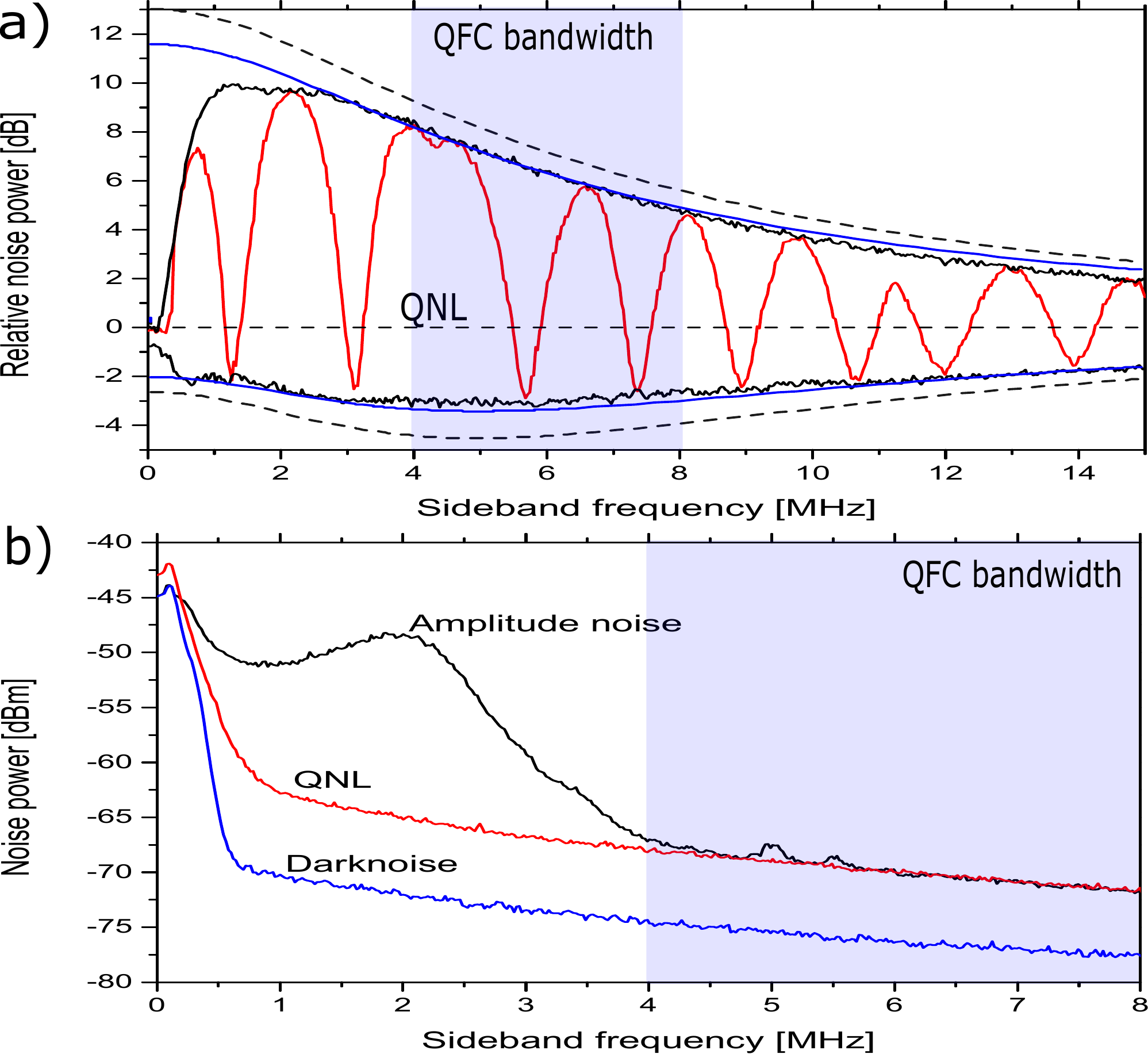}}
\caption{a) The figure shows the relative noise power, while scanning the relative phase between the local oscillator (LO) and the squeezed vacuum beam. The dashed black and solid blue lines show the theoretical model given by Eq.\ref{eq:sqz} taking into account the estimated detection efficiencies of i) 0.84 and ii) 0.92 and taking account for phase fluctuations, which limits the observed squeezing level to -3.5 dB directly detected with our homodyne detector and -4.2~dB just before the SFG cavity. The black experimental data traces are the minimum and maximum noise power measured over 5 minutes with the spectrum analyser.  b) Noise power measured of the up-converted 1064~nm coherent beam to blue. Measured with self-homodyne detection. The blue shaded area shows the QFC bandwidth region which is limited by classical noise of the 1064~nm seed and the SFG cavity bandwidth of 8.7~MHz. The black trace is the amplitude noise (measured with one diode), the red trace is the subtracted photo current and gives the quantum noise limit (QNL) and the blue trace is the electronical dark noise.}
\label{fig2}
\end{figure}

The 1064 nm vacuum squeezed states are generated with an OPA, in which we use a back seeded/locked bow-tie cavity. The lock beam is send through a phase-plate generating a flip-mode and shifted 275 MHz in order to be on resonance with the TEM$_{01}$, while the squeezing direction is on resonance with a vacuum TEM$_{00}$ mode, hereby avoiding any possible destructive interference. The nonlinear crystal used is a $1\times2\times10$~mm$^3$ PPKTP crystal, and the bow-tie cavity consists of two curved mirrors of 38 mm radius of curvature and two plane mirrors. All mirrors are super-polished in order to minimize scattering losses and three of the mirrors are highly reflective at 1064 nm, $R > 99.94~\%$, while the output coupler has a transmission of $T = 12~\%$. The transmittance of the mirrors at the pump wavelength, 532 nm, is more than 95$\%$. The cavity has a free spectral range (FSR) of 767($\pm4$)~MHz, and a finesse of 55($\pm5$), resulting in a cavity bandwidth of 14($\pm0.7$)~MHz, which sets the squeezing bandwidth. The intracavity passive loss were calculated from measurements of the cavity transmission to be $0.3(\pm0.1)\%$.

In Fig.~\ref{fig2}a), the measured 1064 nm vacuum squeezed state is shown in the range from 0 to 15 MHz. The squeezing was measured with a spectrum analyser with a resolution bandwidth of 200 kHz and video bandwidth of 200 Hz. The measured squeezing level was 3.5($\pm0.2$)~dB. However, phase fluctuations dominate the measurement and the full squeezing strength could therefore not be measured. Below 500 kHz the squeezing is limited by the cut-off high-pass frequency of our balanced homodyne detector. Taking into account the losses and phase fluctuations, the actual amount of squeezing available for up-conversion can be estimated using the following relation between squeezing and anti-squeezing \cite{Takeno2007}:  
\begin{equation}
V^{\mp}(\theta)=V^{\mp}\cos^2(\theta)+ V^{\pm}\sin^2(\theta), 
\label{eq:sqz}%
\end{equation}
where $\theta$ is the rms phase fluctuations and the squeezing- and anti-squeezing quadratures are given by
\begin{equation}
V^{\mp} = 1 \mp \eta \frac{4\sqrt{P/P_t}}{4\Omega^2 + (1 \pm \sqrt{P/P_t})^2}. 
\label{eq:sqz2}%
\end{equation}
Here, $\Omega$ is the detection frequency relative to the cavity bandwidth, $\eta$ is the overall quantum detection efficiency and $P/P_t$ is the relative pump power $P$ to the OPA threshold $P_t$. Using a 532-nm SH pump power of 350 mW and scanning the relative phase between the pump beam and the seed beam. Note that the bright seed is only used when estimating the gain. We measure a maximum parametric gain of 7($\pm$2), which gives a pump to threshold ratio of $\sqrt{P/P_t}\approx0.6$. The squeezing is measured using homodyne detection (HD) with a visibility of 97$\%$. The propagation loss from the OPA to the HD is approximately 5$\%$, the quantum efficiencies of the pin-diodes are 97($\pm 2$)$\%$, and the cavity escape efficiency is 97$\%$, resulting in an overall quantum detection efficiency for the squeezing generation of $\eta=84(\pm2)\%$. The electronic dark noise was more than 17 dB below the quantum noise limit (QNL), and was therefore insignificant relative to the observed squeezing levels.

In Fig.~\ref{fig2}a) the blue solid lines are given by Eq.~\ref{eq:sqz}, using the experimental measured parameters and phase fluctuations with an rms of $\theta=0.18$. The black dashed line in Fig.~\ref{fig2}a) takes into account only the phase fluctuations, cavity escape efficiency, and the losses from the OPA to SFG, approximately 5$\%$ loss. Hereby we can estimate the amount of squeezing for the up-conversion to blue light is approximately 4.2($\pm 0.2$)~dB at a sideband frequency of 5 MHz. This inferred squeezing value is just in front of the SFG before the vacuum state is seeded into the SFG.     

\section{The double-resonant SFG}

\begin{figure}[htbp]
\centering
\fbox{\includegraphics[width=0.9\linewidth]{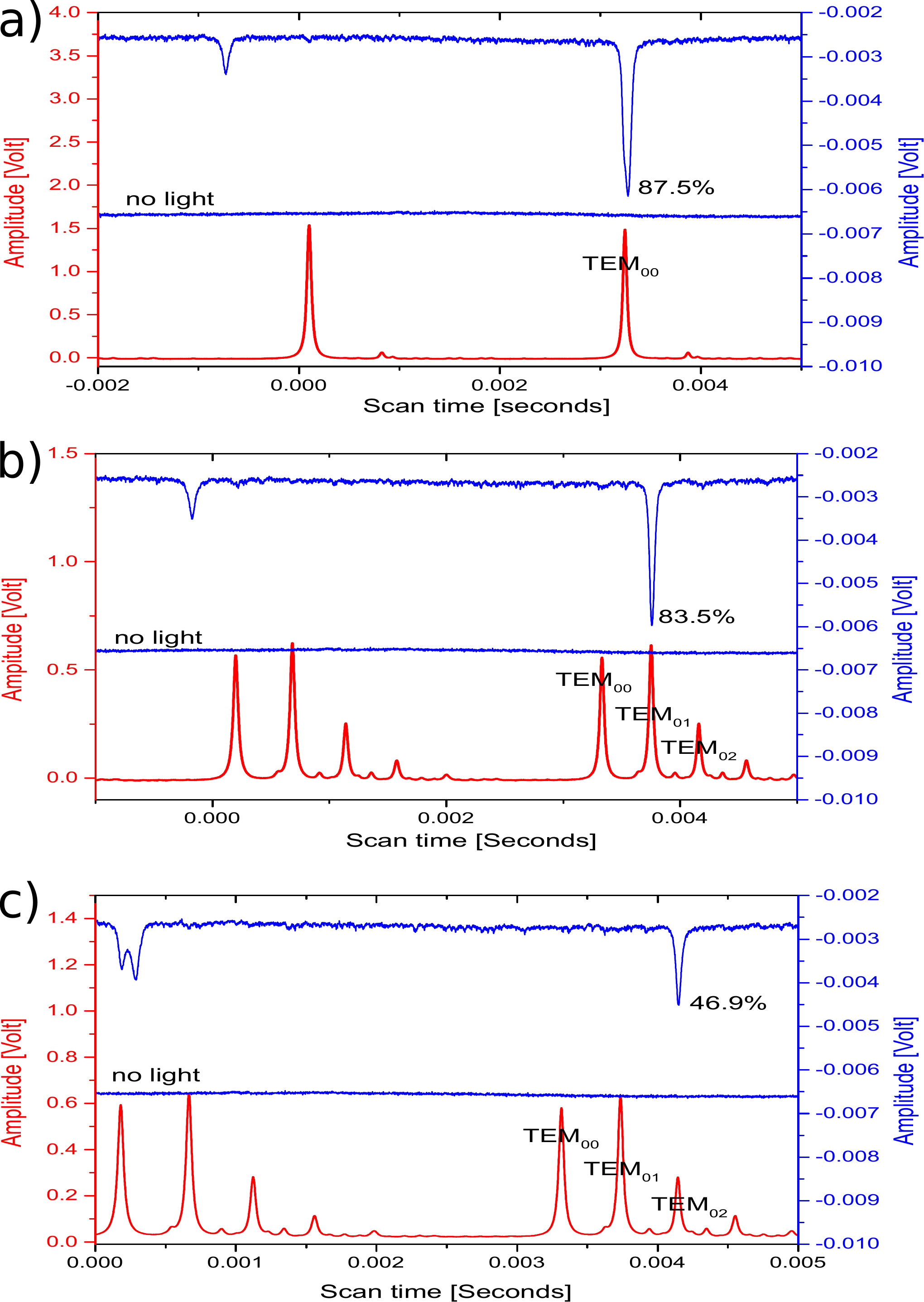}}
\caption{SFG cavity response while scanning the cavity length over one FSR, showing the reflection from the cavity input mirror at 1064 nm (blue trace) and for the 852 nm transmission (red trace) through the cavity and also the impedance matching for the different modes are shown. a) For a TEM$_{00}$ generated SFG mode, b) For a TEM$_{01}$ generated SFG mode, and c) for a TEM$_{02}$ generated SFG mode.}
\label{fig3}
\end{figure}

For the double-resonant SFG we use a cavity which is identical in construction to the one used for the OPA. The SFG cavity, has an input coupler with a transmission of $T = 8\%$ at 1064~nm and 852$\pm 20$~nm while remaining three mirrors are highly reflective at both 1064 nm and 852~nm, $R > 99.94\%$. In order to keep both beams on resonance with the cavity, we apply two PDH locking methods. The first PDH lock controls the cavity length via an intracavity piezo-actuated mirror to keep the cavity on resonance at 1064~nm. The zero order diffracted beam from the AOM (not frequency shifted) is use for locking the length/frequency of the SFG cavity by back seeding it through one of the plain HR windows. The 1064 nm locking beam also acts as a displacement of the incident vacuum squeezing beam for generating a bright blue squeezed beam. The 1064 nm vacuum squeezed beam is 98$\%$ mode-matched into the SFG cavity via the 8$\%$ input coupler.  A total of 330~$\mu$W of 1064 nm seed (displacement) power was transmitted through the SFG cavity. The second PDH lock stabilizes the Ti:sapphire laser frequency in order to maintain the double resonance. This stabilization technique locks the Ti:sapphire laser frequency to the DR SFG cavity, and thus to the 1064-nm laser, ensuring a high frequency stability of the SFG output field. The cavity has an FSR of 767($\pm 4$)~MHz, a finesse of 88($\pm 5$), and thus a cavity bandwidth of 8.7($\pm 0.5$)~MHz. More details about the SFG cavity and the theoretical analysis can be found in our recent published work on double-resonant highly efficient SFG \cite{Kerdoncuff2020}. In Fig.~\ref{fig2}b) we show the noise-power measurements of the generated SFG light (the cavity transfer function of the SFG cavity), when seeded with a 1064 nm coherent state and pumped with 260 mW of optical power from the Ti:sapphire laser, hereby generating 220($\pm$20)~$\mu$W of blue light. From Fig.~\ref{fig2}b) it is seen that our SFG squeezing bandwidth is limited to approximately 4-8 MHz due to classical noise in our 1064 nm seed beam and the 8.7 MHz cavity bandwidth of the SFG cavity. The dark noise of our detector was approximately 5.5 dB below the QNL.

Figure~\ref{fig3} shows the SFG cavity response while scanning the cavity length over one FSR using a 1064 nm TEM$_{00}$ mode and different 850 nm TEM$_{0n}$ pump modes. This is achieved simply by slightly misaligning the 850 nm pump beam into the SFG cavity in order to excite higher-order Hermite-Gaussian (HG) modes. The SFG cavity is then locked to the TEM$_{0n}$ mode. The pump induced increased intracavity loss alters the coupling of the incident squeezed seed field into the cavity, which can go from overcoupled to impedance-matched. We infer the mode-matching of the different modes of the seed beams simply by measuring the drop of reflected power. As seen from Fig.~\ref{fig3} we measure an incoupling efficiency for the squeezed seed beam of 87.5(2)$\%$, 83.5(2)$\%$ and 46.9(2)$\%$, when using a TEM$_{00}$, TEM$_{01}$ and TEM$_{02}$ pump mode, respectively. The intracavity passive loss was also calculated from these measurements of the cavity reflection and mode-matching to be $0.3(\pm 0.1)\%$. When both the 850 nm pump and 1064 nm seed fields are resonant and the phase-matching conditions are fulfilled, 220~$\mu$W, 177~$\mu$W, and 138$\mu$W of blue light from a seed (displacement) power of 330 $\mu$W were generated. The reason for the observed drop in power with increasing HG mode order, is both due to a decreasing nonlinear mode overlap (between the TEM$_{0n}$ pump mode and the TEM$_{00}$ seed mode) and a reduced coupling efficiency into the cavity for the higher-order pump modes \cite{Lassen2006}. We expect that the QFC efficiency for the TEM$_{00}$, TEM$_{01}$, and TEM$_{02}$ mode should be 0.58, 0.45 and 0.20, respectively, hereby resulting in blue bright squeezing levels of 1.9 dB, 1.4 dB and 0.6 dB for the three different modes. The QFC efficiency is estimated by measuring the amount of generated blue light relative to the amount of 1064 nm used in the SFG process.  From Fig.~\ref{fig3}a) it is seen that a very small level of depletion of the 850 nm pump field is observed for pumping with a TEM$_{00}$ mode, but in general the pump field is treated classically in our analysis, thus the noise of the pump laser will not contribute to the squeezing spectrum. In the following we will experimentally investigate the generation of tunable squeezed states in the blue wavelength range. 

\section{Tunable quantum frequency conversion}

\begin{figure}[htbp]
\centering
\fbox{\includegraphics[width=0.9\linewidth]{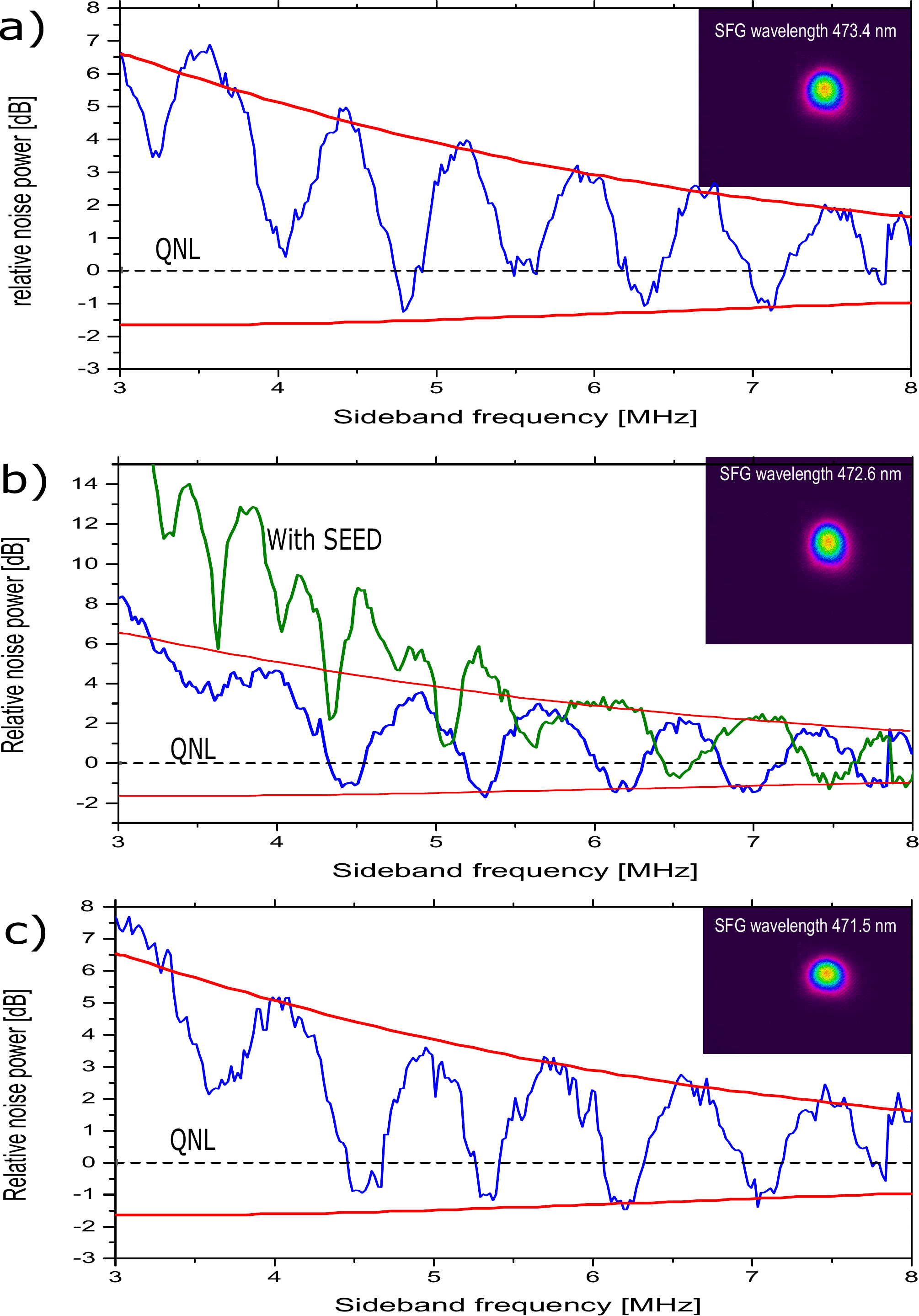}}
\caption{Measurement of relative noise power of the up-converted squeezed vacuum states to bright squeezed states at different wavelengths, a) 473.4 nm, b) 472.6 nm and c) 471.5 nm with approximately 220~$\mu$W of optical power. The measured squeezing level is approximately 1.5~dB. In b) the green trace shows data when using a bright 1064 nm squeezed beam. The red solid lines are the theoretical model (Eq.\ref{eq:sqz} where the function has been multiple with a Lorenz function using 8.7 MHz as bandwidth and a QE of 0.49.}
\label{fig4}
\end{figure}

The measured up-converted squeezed state, while scanning the phase of the OPA 532 nm pump, for different wavelengths are shown in Fig.~\ref{fig4} for a) 473.4 nm, b) 472.6 nm, and c) 471.5 nm. The wavelength tuning of the SFG output is currently limited by the phase-matching of our PPKTP nonlinear crystal to about 473 nm $\pm$~4 nm, although, this can be expanded to cover the entire blue spectrum (420 nm to 510 nm) by proper choice of the second-order nonlinear crystal. The solid red lines in Fig.~\ref{fig4} are plotted using Eq.~\ref{eq:sqz} multiplied by the SFG response/transfer cavity function taking into account the extra losses and the SFG cavity, as described in \cite{Baune2015}, with a cavity bandwidth of 8.7 MHz and a QE of 0.49. We find that the theoretical model is an very good agreement with the measured squeezing spectrum taking into account transfer functions of the SFG cavity and phase fluctuations. The generated SFG beam is in a diffraction limited TEM$_{00}$ mode with M$^2$ < 1.05 \cite{Kerdoncuff2020}. The high spatial-mode quality of the SFG output is readily provided by the DR cavity design and makes our SFG system potentially useful as a light source for various types of microscopes, high spatial resolution scatterometry, and dark-field wafer inspection. The squeezing spectra are measured with self-homodyne detection, thus only the amplitude quadrature is measured, and recorded on a spectrum analyzer with a resolution bandwidth of 200 kHz and video bandwidth of 200 Hz. As explained in the previous section, the SFG up-conversion process is seeded with a 4.2-dB vacuum squeezed state at 1064 nm and converted to a 1.5($\pm$0.2)~dB bright blue squeezed state with approximately 220 $\mu$W of optical power. The reason for not locking our 1064 nm squeezing quadrature to the amplitude squeezing can be seen from Fig.~\ref{fig4}b), where the SFG is seeded with a bright 1064 nm squeezed beam with approximately 55 $\mu$W of optical power. We find that the classical noise of our 1064 nm laser and the limited bandwidth of our SFG cavity is masking the squeezing and makes it impossible to measure the full converted squeezing level. However, having a higher bandwidth of SFG cavity a bright seed might be used and the squeezing quadrature can be locked using a seed. The SFG squeezing data has not been corrected for electronic dark noise. The effect of electronic noise is to reduce the observed amount of squeezing. Thus, by subtracting the electronic noise, the inferred squeezing is approximately 1.6($\pm0.2$)~dB. This gives a total QFC efficiency of 0.49($\pm$0.05) from 1064 nm to bright blue squeezed light. The discrepancy between the calculated and the estimated efficiencies, suggests that additional losses are present in the system. We therefore compared the expected QFC efficiency with the measured, and estimate that the QE of our self-homodyne detector for detecting blue squeezed light is approximately 0.91($\pm$0.05). Note that we can infer about 3.0($\pm$0.2)~dB of tunable squeezing by neglecting the phase-noise and have 99~$\%$ incoupling efficiency and homodyne detectors with unity quantum efficiency. To our knowledge, this is the first demonstration of tunable squeezed light in the blue wavelength range using an SFG stage seeded with 1064 nm vacuum squeezed states.

\section{Quantum frequency conversion to higher-order modes.}

\begin{figure}[htbp]
\centering
\fbox{\includegraphics[width=0.9\linewidth]{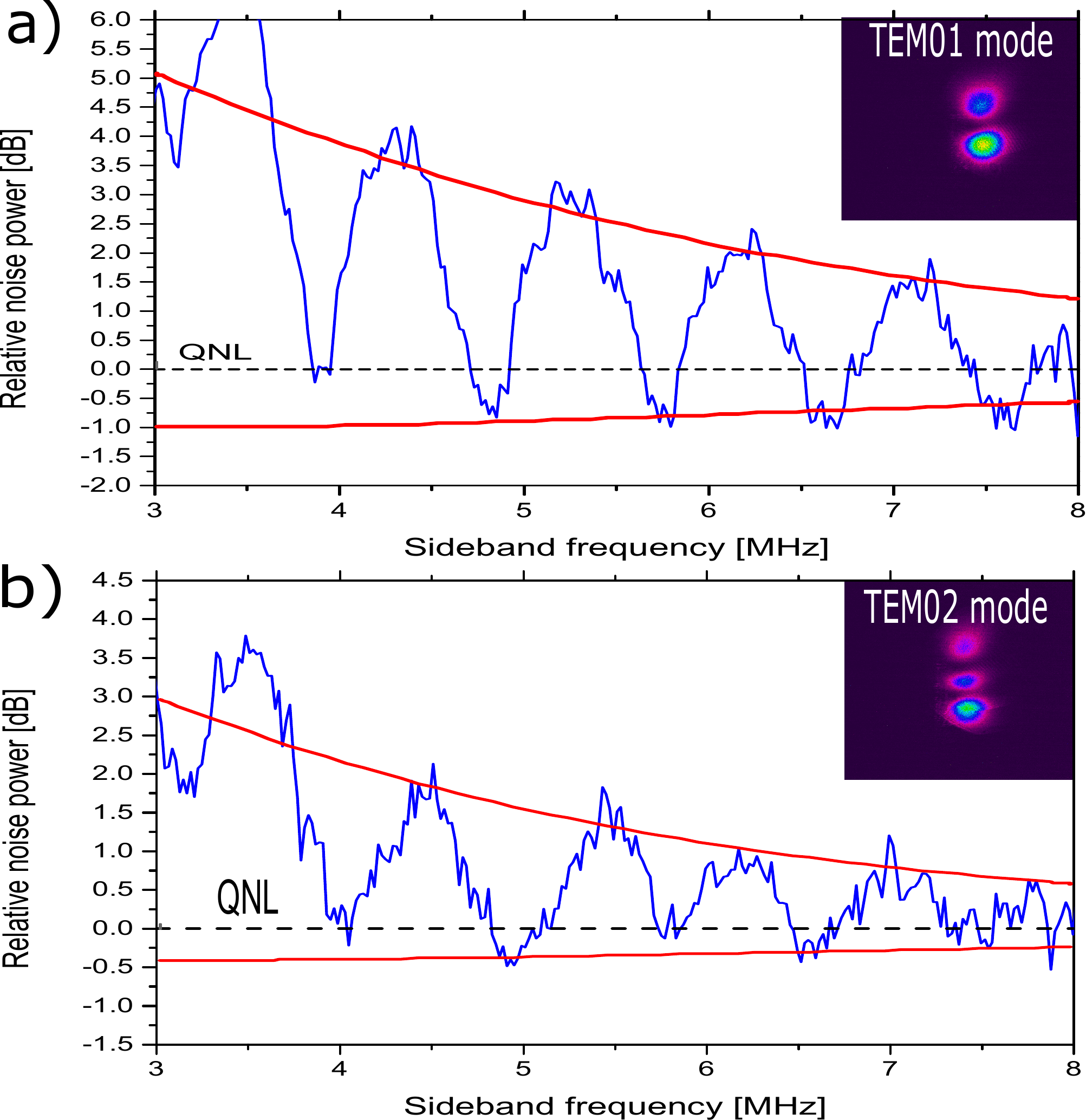}}
\caption{Measurement of the up-converted squeezed vacuum states for different high-order modes a) TEM$_{01}$, and b) TEM$_{02}$ modes, recorded by self-homodyne detection. While scanning the relative phase of the OPA 532 nm pump beam. Resolution bandwidth of 200 kHz and video bandwidth of 200~Hz. The red solid lines are the theoretical model (Eq.\ref{eq:sqz}, where the function again has been multiple with the cavity transfer function using 8.7 MHz as bandwidth and a QE of 0.36 and 0.15 for the TEM$_{01}$ and TEM$_{02}$ modes, respectively. }
\label{fig5}
\end{figure}

In the previous section we demonstrated tunable squeezing of a TEM$_{00}$ mode. We now focus on QFC of spatial higher-order modes. Where there is a growing interest in spatial quantum optical effects, usually called quantum imaging \cite{Lugiato2002}, as the generation of spatial correlations or spatial squeezing in the transverse plane of laser beams may be applied to applications in different areas, such as biophotonics, laser pbridges the gap between a variety of quantum systemshysics, astronomy and quantum information. Here we will concentrate on the two first HG modes, but a similar study could be undergone with another set of modes. Producing squeezed light in any of the TEM$_{nm}$ modes using QFC with SFG requires optimization of different conditions. The SFG pump cavity needs to be resonant for the HG mode of interest. This is achieved simply by misaligning the pump beam in order to optimize coupling into the desired spatial pump mode and then lock the length of the cavity to this resonance. Furthermore the spatial profiles of the pump beam and the seed beam need to be chosen to maximize the spatial overlap between the cavity mode and pump mode and hereby optimizing the nonlinear conversion efficiency of the SFG interaction. The optimized spatial mode pump mode has not been investigated here. However, this can be achieved using a spatial light modulator to synthesize the high-order pump mode \cite{Lassen2006,LassenPRL2007}. In Fig.~\ref{fig5} the measured up-converted squeezed state, while scanning the phase of the OPA SH pump, for two first HG modes a) TEM$_{01}$ and b) TEM$_{02}$ are shown. We seed the SFG with a 4.2($\pm$0.2)~dB vacuum squeezed TEM$_{00}$ mode at 1064 nm and convert it to a 1.1($\pm$0.2)~dB and 0.4($\pm$0.2)~dB blue squeezed state in the TEM$_{01}$ and TEM$_{02}$ modes, receptively. The reason for the lower levels of squeezing compared to the TEM$_{00}$ mode is due to a lower level of incoupled pump power, thus a decrease in the nonlinear mode-overlap. However, having higher levels of incoupled pump power available similar squeezing levels as for the TEM$_{00}$ mode should be achievable. We estimate that the mode-overlap between TEM$_{00}$ mode and TEM$_{01}$ mode and TEM$_{02}$ mode is 0.58 and 0.26, respectively. Resulting in that the power levels should be 2 and 4 times larger compared to the TEM$_{00}$-TEM$_{00}$ mode coupling. From the measured squeezing we estimated that the total QFC efficiency is 0.36($\pm$0.05) and 0.15($\pm$0.05) for the TEM$_{01}$ and TEM$_{02}$ modes, receptively. We would like to note that this, to our best knowledge, is the first demonstration of bright higher-order transverse mode squeezing using QFC.

\section{Conclusion}
%%%%%%%%%%%%%%%%%%%%%%%
We have demonstrated quantum frequency conversion of 4.2($\pm$0.2) dB vacuum squeezed states at 1064 nm into the blue wavelength range using a double-resonant sum-frequency conversion cavity. We demonstrate noise squeezing levels of 1.6($\pm$0.2)~dB relative to the quantum noise limit with an overall conversion/detection efficiency of 0.49($\pm$0.05). The wavelength tuning of the SFG output is limited by the phase-matching of the nonlinear crystal to about 8 nm. However, this can be expanded to cover the entire blue spectrum (420 nm to 510 nm) by proper choice of second-order nonlinear crystal. Currently phase noise fluctuations limit us, however, we can infer 3.0($\pm$0.2)~dB of tunable squeezing by neglecting the phase-noise and assuming 99$\%$ incoupling efficiency and homodyne detection with a quantum efficiency of unity. 3 dB of squeezing is often considered the benchmark level for real-life quantum applications. For example for quantum sensing this value provides a signal to shot noise enhancement that corresponds to doubling the light power compared to using quantum noise limited light. Measurements of the beam profile of the SFG light shows a diffraction limited TEM$_{00}$ mode with M$^2$ < 1.05. We foresee that the tunabillity and the high spatial quality of the SFG output makes our light source useful as a source for various types of microscope and spectroscopic measurements. The SFG system can be adapted for applications for life science and biomedical research, where various fluorophores are known to excite close to a wavelength of 488 nm. Bright squeezing in high-order spatial modes was demonstrated by generating squeezing levels of 1.1($\pm$0.2)~dB and 0.4($\pm$0.2)~dB in the TEM$_{01}$ and TEM$_{02}$ modes, receptively. Higher-order spatial mode squeezed light has promising application for super resolution quantum imaging, mode matching loss in interferometry, and realizing CV parallel quantum information protocols based on spatial multimode squeezed states. We plan for future experiments to produce spatial entanglement between modes with orbital angular momentum using a spatial light modulator to synthesize any type of multi-mode pump mode and to investigate super resolution quantum imaging.

%%%%%%%%%%%%%%%%%%%
\section*{Funding}
The Danish Agency for Institutions and Educational Grants. The Eureka turbo project "Quantum-Gravity Wave Detection" (EUROS E11677 T-Q-GWD) and the EMPIR-EURAMET project (17FUN01 "BeCOMe"). 

%%%%%%%%%%%%%%%%%%%%%%%%%%%%
\section*{Acknowledgments}
We acknowledge Jan Thomsen, NBI, for lending us the Ti:sapphire laser and Tobias Gehring, DTU, for lending us the diodes used for the homodyne detectors.   

%%%%%%%%%%%%%%%%%%%%%%%%%%%%
\section*{Disclosures}
The authors declare no conflicts of interest.

%%%%%%%%%%%%%%%%%%%%%%%%%%%%
\section*{Data availability} Data underlying the results presented in this paper are not publicly available at this time but may be obtained from the authors upon reasonable request.

%%%%%%%%%%%%%%%%%%%%%%% References %%%%%%%%%%%%%%%%%%%%%%%%%

\end{document}